\documentstyle[12pt]{article}

\begin{document}
\begin{titlepage}

\centerline{\bf GENERAL RELATIVITY ON A NUL SURFACE:} 
\centerline{\bf HAMILTONIAN FORMULATION IN THE TELEPARALLEL GEOMETRY}
\vskip 1.0cm   
\bigskip
\centerline{\it J. W. Maluf$\,^{*}$ and J. F. da Rocha-Neto}
\centerline{\it Departamento de F\'isica}
\centerline{\it Universidade de Bras\'ilia}
\centerline{\it C.P. 04385}
\centerline{\it 70.919-970  Bras\'ilia, DF}  
\centerline{\it Brazil}
\date{}
\begin{abstract}
The Hamiltonian formulation of general relativity on a null surface is
established in the teleparallel geometry. No particular conditions on
the tetrads are imposed, such as the time gauge condition. By means of
a 3+1 decomposition the resulting Hamiltonian arises as a 
completely constrained system. However, it is
structurally different from the standard Arnowitt-Deser-Misner (ADM) 
type formulation. In this geometrical framework the basic field
quantities are tetrads that transform under the global SO(3,1)
and the torsion tensor.

\end{abstract}
\thispagestyle{empty}
\vfill
\par
\noindent (*) e-mail: wadih@fis.unb.br
\end{titlepage}
\newpage

\noindent {\bf I. Introduction}\par
\bigskip
\noindent The study of asymptotically flat gravitational waves is 
an important and interesting issue in general relativity. It started
with the pioneering work of Bondi\cite{Bondi}, which was
subsequently generalized by Sachs\cite{Sachs}. It was soon realized
that the description of gravitational waves on a null surface 
facilitates the characterization of the true, independent
degrees of freedom of the gravitational field. This
characterization may possibly be mandatory to the 
quantization of gravity. Moreover, null surfaces
play an important role in the study of gravitational radiation.

Difficulties in working with the dynamics of null surfaces are well
known. The latter are characterized by the condition $g^{00}=0$. 
However, if we naively impose this condition in Einstein's equations
we spoil the six evolution equations, since these equations become 
exempt of second order time derivatives and consequently the 
evolution becomes undetermined. Imposition of the above condition 
in the variation of the Hilbert-Einstein action integral leads to 
nine equations only. Therefore attempts have been made
to arrive at a well posed {\it characteristic} initial value problem.

The analysis of the initial value problem for asymptotically flat,
nonradiating space-times is reasonably well understood. Moreover,
the Arnowitt-Deser-Misner (ADM) Hamiltonian formulation\cite{ADM} 
is usually taken as a paradigm to the study of the dynamics 
of spacelike surfaces. In contrast, there does not seem to exist
a widely accepted formulation of the characteristic initial value
problem, or of the corresponding Hamiltonian formulation, as we
observe from the vast literature on the subject. The initial value
problem has been analysed, for instance, in 
\cite{Inverno1,Inverno2,Inverno3,Inverno4,Brady}, whereas the 
Hamiltonian formulation has been investigated both in a 2+2 
decomposition (\cite{Torre,Inverno5}) and in a 3+1 
decomposition (\cite{Goldberg1,Goldberg2,Goldberg3,Goldberg4}).
In particular, the work of refs.\cite{Inverno5,Goldberg3,Goldberg4}
is developed in the context of Ashtekar variables. While all of these
approaches add some progress to the understanding of the dynamics 
of the gravitational field on null surafaces, we see
that at the present time there does not exist a definite,
irrefutable Hamiltonian formulation which would, accodding to
Goldberg {\it et. al.}\cite{Goldberg3}, 
display in an isolated form the true 
degrees of freedom and the observables of the theory, in such a 
way that the dynamics of these degrees of freedom is singled out
from the dynamics of the remaining field quantities.

In this paper we construct the Hamiltonian for the gravitational
field on a null surface in the teleparallel geometry. The 
analysis of the dynamics of spacelike surfaces 
in this geometry has already been  carried out in \cite{Maluf1}. 
However in that analysis the time gauge
condition was imposed in order to simplify the considerations.
Since we cannot impose at the same time the null surface and the 
time gauge conditions, the problem has to reconsidered in a new
fashion.

The analysis of the gravitational field in this 
geometrical framework has 
proven to be useful, among other reasons because of the appearance 
of a scalar density in the form of a divergence in the Hamiltonian 
constraint, and  which is identified as the gravitational energy 
density\cite{Maluf2}. This expression for the gravitational energy
can be applied to concrete, physical configurations of the 
gravitational field (see, for instance, 
refs.\cite{Maluf2,Maluf3,Maluf4,Maluf5}).
In this paper we obtain a similar structure. The four constraints
of the theory contain each one a divergence which altogether
constitute a vector density, and
which strongly suggests a definition for the
gravitational radiation energy. The
detailed analysis of this issue is not carried out here.

One achievement of this long term program is to demonstrate
that general relativity can be alternatively presented and discussed
in the teleparallel geometry, without recourse to the Riemann
curvature tensor or to the Levi-Civita (metric compatible) 
connection. In this sense, this geometrical framework  
allows an alternative understanding of the gravitational field.

In section II we present the Lagrangian formulation of the teleparallel
equivalent of general relativity (TEGR) in a way somewhat different 
from what has been presented so far. In \cite{Maluf1} the theory
is formulated initially with a local SO(3,1) symmetry, and in the
Hamiltonian analysis, after fixing the time gauge condition, 
it is concluded that in order to arrive at a
set of first class constraints it is necessary to transform the 
SO(3,1) into a global symmetry group. In this paper the symmetry group 
is taken as the global SO(3,1) from the outset. In section III we  
present the boundary conditions for the tetrad 
components, assuming that the radiation is due to a localized source.
In section IV we present in detail the construction of
the Hamiltonian, obtained by a 3+1 decomposition.
In the last section we present additional comments and point out 
further developments.\par
\bigskip
\noindent Notation: spacetime indices $\mu, \nu, ...$ and SO(3,1)
indices $a, b, ...$ run from 0 to 3. In the 3+1 decomposition latin 
indices from the middle of the alphabet indicate space indices according 
to $\mu=0,i,\;\;a=(0),(i)$. The tetrad field $e^a\,_\mu$ 
yields the usual definition of the torsion tensor:  
$T^a\,_{\mu \nu}=\partial_\mu e^a\,_\nu-\partial_\nu e^a\,_\mu$.
The flat, Minkowski spacetime  metric is fixed by
$\eta_{ab}=e_{a\mu} e_{b\mu}g^{\mu\nu}= (-+++)$.        \\

\bigskip
\bigskip
\noindent {\bf II. The Lagrangian formulation of the TEGR}\par
\bigskip

\noindent In \cite{Maluf1} the Lagrangian formulation of the 
TEGR is presented in terms of the tetrad field and a spin connection
$\omega_{\mu ab}$. Both quantities transform under the local
SO(3,1) group but are not related, not even by the field equations.
The equivalence of the teleparallel Lagrangian with
the Hilbert-Einstein Lagrangian holds provided we require the 
vanishing of the curvature tensor tensor $R^a\,_{b \mu \nu}(\omega)$.
In the Hamiltonian analysis we conclude that the symmetry group
must be the global SO(3), and eventually the connection is 
discarded.

In this paper we will establish the Lagrangian density in terms of 
the tetrad field only. The symmetry group is the global SO(3,1). 
The Lagrangian density is given by

$$L(e)\;=\;-k\,e\,\Sigma^{abc}T_{abc}\;,\eqno(1)$$

\noindent where $k={1\over {16\pi G}}$, $G$ is Newton's constant,
$e=det(e^a\,_\mu)$, $T_{abc}=e_b\,^\mu e_c\,^\nu T_{a \mu \nu}$ and

$$\Sigma^{abc}\;=\;{1\over 4}(T^{abc}+T^{bac}-T^{cab})+
{1\over 2}(\eta^{ac}T^b-\eta^{ab}T^c)\;.\eqno(2)$$

\noindent Tetrads transform space-time into SO(3,1) indices
and vice-versa. The trace of the torsion tensor is given by 

$$T_b=T^a\,_{ab}\;.$$

\noindent The tensor $\Sigma^{abc}$ is defined such that

$$\Sigma^{abc}T_{abc}\;=\; {1\over 4} T^{abc}T_{abc} + 
{1\over 2}T^{abc}T_{bac}-T^aT_a\;.$$

The field equations obtained from (1) read

$${{\delta L}\over {\delta e^{a\mu}}}\;=\;
e_{a\lambda}e_{b\mu}\partial_\nu(e\Sigma^{b\lambda \nu})-
e\biggl(\Sigma^{b \nu}\,_aT_{b\nu \mu}-
{1\over 4}e_{a\mu}T_{bcd}\Sigma^{bcd}\biggr)\;=\;0\;.\eqno(3)$$

\noindent It can be shown by explicit calculations\cite{Maluf1} 
that these equations yield Einstein's equations:

$${{\delta L}\over {\delta e^{a\mu}}}\; \equiv \;{1\over 2}\,e\,
\biggl\{ R_{a\mu}(e)-{1\over 2}e_{a\mu}R(e)\biggr\}\;.$$

In order to obtain the canonical formulation we need a first order
differential formulation of (1). This is easily obtanained 
through the introduction of  an  auxiliary field quantity
$\phi_{abc}=-\phi_{acb}$. The first order differential formulation
of the TEGR is described by the following Lagrangian density,

$$L(e,\phi)\;=\;k\,e\,\Lambda^{abc}(\phi_{abc}\,-\,2T_{abc})\;,\eqno(4)$$

\noindent where $\Lambda^{abc}$ is defined in terms of $\phi^{abc}$ 
exactly as $\Sigma^{abc}$ in terms of $T^{abc}$:

$$\Lambda^{abc}\;=\;{1\over 4}(\phi^{abc}+\phi^{bac}-\phi^{cab})+
{1\over 2}(\eta^{ac}\phi^b-\eta^{ab}\phi^c)\;.\eqno(5)$$

\noindent Variation of the action constructed out of (4) with respect
to $\phi^{abc}$ yields

$$\Lambda_{abc}\;=\;\Sigma_{abc}\;,\eqno(6)$$

\noindent which, after some manipulations, can be reduced to

$$\phi_{abc}\;=\;T_{abc}\;.\eqno(7)$$

\noindent The equation above may be split into two equations:

$$\phi_{a0k}\;
=\;T_{a0k}\;=\;\partial_0 e_{ak}-\partial_k e_{a0}\;,\eqno(8a)$$

$$\phi_{aik}\;
=\;T_{aik}\;=\;\partial_i e_{ak}-\partial_k e_{ai}\;.\eqno(8b)$$

\noindent Taking into account eq. (7), it can be shown that the
second field equation, the
variation of the action integral with respect to 
$e_{a\mu}$, leads precisely to (3). Therefore (1) and (4) exhibit
the same physical content.

In section IV we will make explicit reference to null surfaces. The
theory defined by (1) or (4) describes an arbitrary gravitational
field, as there is no restriction in the form of a Lagrange 
multiplier fixing some particular geometry. Without going into 
details we just mention that if we impose the condition
$g^{00}=0$ in (3) the resulting equation will still have
second order time derivatives
(note that this equation has one SO(3,1) and one space-time index).

Before closing this section let us make a remark. The theory defined
by (1) and (2) has been considered in the literature, in a different 
context, as the translational gauge formulation of Einstein's general
relativity\cite{Cho}. It is argued in this approach
that (1) is invariant under
{\it local} SO(3,1) transformations up to a total divergence. This
divergence is then discarded, from what is concluded that (1) 
exhibits local gauge symmetry. We do not endorse this point of view. 
A careful analysis of this divergence
(the last term of eq. (12) of \cite{Cho}) shows that in general it 
does not vanish for arbitrary elements of the SO(3,1) group when
integrated over the whole spacelike surface. Problems arise if
the SO(3,1) group elements fall off as $const.\,+\,O({1\over r})$ when
$r\rightarrow \infty$. Therefore the action is
not, in general, invariant under such transformations. 
Surface terms play a very important role in action integrals for the
gravitational field, so that one cannot {\it arbitrarily} add or remove 
them. Moreover, if (1) were actually invariant under the local
SO(3,1) group, then the theory would have six additional
constraints, which would spoil the counting of degrees of freedom
of the theory (see eqs. (18), (19) and (31) ahead).  \\
\bigskip
\bigskip

\noindent {\bf III. The boundary conditions}\par
\bigskip

In order to guarantee that the space-time of a localized radiating
source is asymptotically flat we adopt the conditions laid
down by Bondi\cite{Bondi} and Sachs\cite{Sachs} for the metric
tensor. The conditions on the tetrads are simply obtained by
constructing the tetrads associated with these radiating fields
and identifying the asymptotic behaviour when $r\rightarrow \infty$.
Of course there is an infinity of tetrads that yield the same 
metric tensor. However, we will consider a typical configuration
and assume the generality of our considerations.
For simplicity we will consider in detail Bondi's metric.

Bondi's metric is not an exact solution of Einstein's equations.
In terms of radiation coordinates $(u,r,\theta,\phi)$, where $u$
is the retarded time and $r$ is a luminosity distance, 
Bondi's radiating metric is written as

$$ds^2\;=\;-\biggl( {V\over r} e^{2\beta}- U^2\,r^2 e^{2\gamma}\biggr)du^2
-2e^{2\beta}du\,dr - 2U\,r^2\,e^{2\gamma}du\,d\theta$$

$$+r^2 \biggl( e^{2\gamma}\,d\theta^2 + 
e^{-2\gamma}\,sin^2\theta\,d\phi^2\biggr)\;.\eqno(9)$$

\noindent The metric above is such that the surfaces
for which $u=constant$ are null hypersurfaces. Each null radial (light)
ray is labelled by particular values of $u, \theta$ and $\phi$. At
spacelike infinity $u$ takes the standard form $u=t-r$. The four 
quantities appearing in (9), $V, U, \beta$ and $\gamma$ are functions of
$u, r$ and $\theta$. Thus (9) displays axial symmetry. A more general
form of this metric has been given by Sachs\cite{Sachs}, who  showed 
that the most general metric tensor describing asymptotically flat 
gravitational waves depends on six functions of the coordinates.

The functions in (9) satisfy the following asymptotic behaviour:

$$\beta\;=\;-{c^2\over {4r^2}}+...$$

$$\gamma\;=\;{ c \over r}+...$$

$${V\over r}\;=\;1\,-\,{{2M}\over r}\,
-\,{1\over r^2}\biggl[ {{\partial d}\over {\partial \theta}} +
d\,cot\theta-\biggl({{\partial c}\over{\partial \theta}}\biggr)^2
-4c\biggl({{\partial c}\over {\partial \theta}}\biggr)cot\theta-
{1\over 2}\,c^2\biggl(1+8cot^2\theta\biggr)\biggr]+...$$

$$U\;=\;{1\over r^2}\biggl( {{\partial c}\over {\partial \theta}}
+2c\,cot\theta\biggr)+ {1\over r^3} \biggl( 2d
+3c\,{{\partial c}\over{\partial \theta}}\,cot\theta+
4c^2\,cot\theta\biggr)+...\;,$$

\noindent where $M=M(u,\theta)$ and $d=d(u,\theta)$ are the mass aspect
and the dipole aspect, respectively, and from the function $c(u,\theta)$
we define the news function ${{\partial c(u,\theta)} \over {\partial u}}$.

One possible realization of this metric tensor in terms of tetrad 
fields is given by

$$e_{a\mu}=\pmatrix{-e^\beta ({V\over r})^{1\over 2}&
-e^\beta( {V\over r})^{-{1\over 2}}&0&0\cr
-r\,U\,e^\gamma\,cos\theta\,cos\phi& 
e^\beta({V\over r})^{-{1\over 2}}sin\theta\,cos\phi&
r\,e^\gamma\,cos\theta\,cos\phi& -r\,e^{-\gamma}\,sin\theta\,sin\phi\cr
-r\,U\,e^\gamma\,cos\theta\,sin\phi& 
e^\beta\,({V\over r})^{-{1\over 2}}sin\theta\,sin\phi&
r\,e^\gamma\,cos\theta\,sin\phi& r\,e^{-\gamma}\,sin\theta\,cos\phi\cr
r\,U\,e^\gamma\,sin\theta& 
e^\beta\,({V\over r})^{-{1\over 2}}cos\theta&
-r\,e^\gamma\,sin\theta & 0\cr}\;.\eqno(10)$$

From the expression above we obtain the asymptotic behaviour
of the tetrad components in cartesian coordinates:

$$e_{(0)0}\;\sim\; 1\,+\,O({1\over r})\,+\,...\;,\eqno(11a)$$

$$e_{(0)k}\;\sim\; 1\,+\,O({1\over r})\,+\,...\;,\eqno(11b)$$

$$e_{(i)0}\;\sim\; O({1\over r})\,+\,...\;,\eqno(11c)$$

$$e_{(i)k}\;\sim\; \delta_{ik} \,
+\,{1\over 2}h_{ik}({1\over r})\,+\,...\;.\eqno(11d)$$

These expressions establish the boundary conditions for the tetrads.
As a final comment, we remark that if we make $M=d=0$ in (9), 
Bondi's metric reduces to the flat space-time metric in radiation
coordinates, and so does expression (10) for the tetrads. It
can be shown that in this case all components of the torsion 
tensor vanish.                                          \\

\bigskip 
\bigskip
\noindent {\bf IV. The 3+1 decomposition}\par
\bigskip
\noindent There are several fundamental differences 
between the analysis of this section and the approach of 
Goldberg {\it et. al.}\cite{Goldberg3,Goldberg4}. In the latter,
complex valued field variables and an
orthonormal set of null vectors adapted to a null surface are 
employed. In contrast, we adopt ordinary, real valued tetrads.
Nevertheless, the present analysis is conceptually the same as that
developed in \cite{Goldberg3,Goldberg4}. We conclude,
however, that it is unecessary to establish a  3+1
decomposition for the tetrads, as it is normally done. 
The Hamiltonian formulation arises naturally in terms of the
four dimensional tetrad field and its inverse, as we will see.

The Hamiltonian formulation is established from the 
first order differential Lagrangian density (4). Space and time
derivatives appear only in $T_{abc}$. Expression (4) can be
rewritten as

$$L(e,\phi)\;=\;-4ke\,\Lambda^{a0k}\,\dot e_{ak}+
4ke\,\Lambda^{a0k}\,\partial_k e_{a0} -2ke\,\Lambda^{aij}\,T_{aij}
+ke\Lambda^{abc}\,\phi_{abc}\;,\eqno(12)$$

\noindent where the dot indicates time derivative, and

$$\Lambda^{a0k}\;=\;\Lambda^{abc}\,e_b\,^0\,e_c\,^k\;,$$

$$\Lambda^{aij}\;=\;\Lambda^{abc}\,e_b\,^i\,e_c\,^j\;.$$

\noindent Thus the momentum canonically conjugated to $e_{ak}$ is 
given by

$$\Pi^{ak}\;=\;-4k\,e\,\Lambda^{a0k}\;,\eqno(13)$$

\noindent Expression (12) is then rewritten as

$$L\;=\; \Pi^{ak}\,\dot e_{ak}-\Pi^{ak}\,\partial_k e_{a0}-
2ke\,\Lambda^{aij}\,T_{aij}+ke\,\Lambda^{abc}\,\phi_{abc}\;.\eqno(14)$$

In order to establish the  Hamiltonian formulation we need to
rewrite the expression above in terms of $e_{ak}, \Pi^{ak}$ and
further nondynamical quantities. However this is not a trivial
procedure. In \cite{Maluf1} the 3+1 decomposition of the theory
was possible, to a large extent because of the time gauge condition
$e_{(i)}\,^0=e^{(0)}\,_k=0$. This condition resulted in a tremendous
simplification of the analysis. It is clear that we cannot 
impose simultaneously the time gauge condition and the null
surface condition. Therefore the present analysis will be 
totally different from that of \cite{Maluf1}.

The construction can be formally carried out in two steps. First,
we substitute the Lagrangian field equation (8b) into (14), so
that half of the auxiliary fields, $\phi_{aij}$, are eliminated
from the Lagrangian. Second, we should be able to express the
remaining auxiliary fields, $\phi_{a0k}$, in terms of the 
momenta $\Pi^{ak}$. This is a nontrivial step.

We need to work out the explicit form of $\Pi^{ak}$. It is 
given by

$$\Pi^{ak}\;=\;k\,e\biggl\{ 
g^{00}(
-g^{kj}\phi^a\,_{0j}-e^{aj}\phi^k\,_{0j}+2e^{ak}\phi^j\,_{0j})$$

$$+g^{0k}(g^{0j}\phi^a\,_{0j}+e^{aj}\phi^0\,_{0j})
\,+e^{a0}(g^{0j}\phi^k\,_{0j}+g^{kj}\phi^0\,_{0j})
-2(e^{a0}g^{0k}\phi^j\,_{0j}+e^{ak}g^{0j}\phi^0\,_{0j})$$

$$-g^{0i}g^{kj}\phi^a\,_{ij}+e^{ai}(g^{0j}\phi^k\,_{ij}-
g^{kj}\phi^0\,_{ij})+2(g^{0i}e^{ak}-g^{ik}e^{a0})
\phi^j\,_{ij} \biggr\}\;.\eqno(15)$$

From now on we  impose the null surface condition

$$g^{00}\;=\;0\;.$$

\noindent The imposition of this condition at the end of the Legendre
transform would render infinities.
Denoting $(..)$ and $[..]$ as the symmetric and 
anti-symmetric parts  of field quantities, respectively, we can
decompose $\Pi^{ak}$ into irreducible components:

$$\Pi^{ak}\;=\;e^a\,_i\,\Pi^{(ik)}+e^a\,_i\,\Pi^{[ik]}+
e^a\,_0\,\Pi^{0k}\;,\eqno(16)$$

\noindent where

$$\Pi^{(ik)}\;=\;k\,e\biggl\{
g^{0k}( g^{0j}\phi^i\,_{0j}+
g^{ij}\phi^0\,_{0j}-g^{0i}\phi^j\,_{0j})$$

$$+g^{0i}(g^{0j}\phi^k\,_{0j}+
g^{kj}\phi^0\,_{0j}-g^{0k}\phi^j\,_{0j})
-2g^{ik}\,g^{0j}\phi^0\,_{0j}\;+\;\Delta^{ik}\biggr\}\;,\eqno(17a)$$

$$\Delta^{ik}\;=\;-g^{0m}(
g^{kj}\phi^i\,_{mj}+g^{ij}\phi^k\,_{mj}-2g^{ik}\phi^j\,_{mj})
-(g^{km}g^{0i}+g^{im}g^{0k}) \phi^j\,_{mj}\;,\eqno(17b)$$

$$\Pi^{[ik]}\;=\;k\,e\biggl\{ -g^{im}g^{kj}\phi^0\,_{mj}+
(g^{im}g^{0k}-g^{km}g^{0i})\phi^j\,_{mj}\biggr\}
\;\equiv \;k\,e\,p^{ik}\;,\eqno(18)$$

$$\Pi^{0k}\; =\;-2k\,e\, (
g^{kj}g^{0i}\phi^0\,_{ij}-g^{0k}g^{0i}\phi^j\,_{ij} )
\;\equiv \;k\,e\,p^k \;.\eqno(19)$$

The crucial observation of this analysis is that only
$\Pi^{(ik)}$ depends on the ``velocities'' $\phi^a\,_{0j}$. 
$\Pi^{[ik]}$ and $\Pi^{0k}$ depend solely on $\phi^a\,_{ij}=T^a\,_{ij}$.
Therefore we can express only six of the ``velocity''  fields
$\phi^a\,_{0j}$ in terms of the momenta $\Pi^{(ik)}$. In order
to find out which components of $\phi^a\,_{0j}$ can be inverted
we decompose the latter identically as

$$\phi^a\,_{0j}\;=\;e^{ai}\,\psi_{ij}+e^{ai}\,\sigma_{ij}
+e^{a0}\,\lambda_j\;,\eqno(20)$$

\noindent with the following definitions:

$$\psi_{ij}\;=\;\psi_{ji}\;=\;
{1\over 2}(\phi_{i0j}+\phi_{j0i})\;,$$

$$\sigma_{ij}\;=\;-\sigma_{ji}\;=\;
{1\over 2}(\phi_{i0j}-\phi_{j0i})\;,$$

$$\lambda_j\;=\;\phi_{00j}$$

\noindent Substituting (20) in (17a) we find that $\Pi^{(ik)}$
depends only on $\psi_{ij}$:

$$\Pi^{(ik)}\;=\;
k\,e\biggl\{2(\,g^{0k}g^{im}g^{0j}\psi_{mj}+
g^{0i}g^{km}g^{0j}\psi_{mj}-g^{0i}g^{0k}g^{mn}\psi_{mn}-
g^{ik}g^{0m}g^{0n}\psi_{mn})$$

$$+\Delta^{ik}\biggr\} \;.\eqno(21)$$

\noindent Therefore if terms like $\sigma_{ij}$ and $\lambda_j$ 
appear in $L$, other than in $\Pi^{ak} \dot e_{ak}$, we would have  
difficulties in performing the Legendre transform, because
they cannot be transformed into any momenta
($\Pi^{[ik]}$ and $\Pi^{0k}$ do not depend on them). Fortunately,
they do not appear.
Let us rewrite $L$ given by (14) in terms of (15) and (20),
assuming from now on that $\phi^a\,_{ij}=T^a\,_{ij}$. It is given by

$$L\;=\;\Pi^{ak}\dot e_{ak}+e_{a0}\,\partial_k \Pi^{ak}
-\partial_k(e_{a0}\Pi^{ak})$$

$$+k\,e(-{1\over 4}g^{im}g^{nj}T^a\,_{mn}T_{aij}-
{1\over 2}g^{jn}T^i\,_{mn}T^m\,_{ij}+
g^{ik}T^j\,_{ji}T^n\,_{nk})$$

$$-{1\over 2}\phi_{a0k}  \biggl\{ \Pi^{ak}+ ke[g^{0i}g^{jk}T^a\,_{ij}
-e^{ai}(g^{0j}T^k\,_{ij}-g^{jk}T^0\,_{ij})
-2(e^{ak}g^{0i}-e^{a0}g^{ki})T^j\,_{ij}]   \biggr\}\;.\eqno(22)$$

\noindent The field $\phi_{a0k}$ appears only in the last line
of the expression above. The terms that appear together with
$\Pi^{ak}$ in this line exactly subtract the last line of (15).
It is possible to check by explicit calculations that the last
line of (22) can be written as

$$-{1\over 2} \psi_{ik}\,(\Pi^{(ik)}-k\,e\,\Delta^{ik})\;.\eqno(23)$$

We can then proceed and complete the Legendre transform. In the
present case the latter amounts to
expressing $\psi_{ik}$ in terms of $\Pi^{(ij)}$. The inversion can
be made and leads to

$$\psi_{ik}\;=\;{1\over 2}\biggl\{
g_{0m}(g_{0i}g_{jk}+g_{0k}g_{ji})P^{mj}
-{1\over 2}(g_{0i}g_{0k}g_{mn}+
g_{0m}g_{0n}g_{ik})P^{mn}\biggr\}\;,\eqno(24)$$

\noindent where

$$P^{ik}\;=\;{1\over {ke}}\Pi^{(ik)}-\Delta^{ik}\;.\eqno(25)$$

Substituting (23) and (24) back in (22) we finally arrive at 
the primary Hamiltonian $H_0=p\dot q -L$:

$$H_0\;=\;-e_{a0}\,\partial_k \Pi^{ak}+
k\,e\,(\, {1\over 4}g^{im}g^{nj}T^a\,_{mn}T_{aij}+
{1\over 2}g^{jn}T^i\,_{mn}T^m\,_{ij}-g^{ik}T^j\,_{ji}T^n\,_{nk}\,)$$

$$+\,ke{1\over 2}(g_{0i}g_{0m}g_{nk}-
{1\over 2}g_{0i}g_{0k}g_{mn})P^{mn}P^{ik}\;.\eqno(26)$$

Since equations (18) and (19) constitute primary constraints, they
have to be added to $H_0$, and so the Hamiltonian becomes

$$H\;=\;H_0\;+\;\alpha_{ik}\,(\Pi^{[ik]}-k\,e\,p^{ik})
\;+\;\beta_k\,(\Pi^{0k}-k\,e\,p^k)\;+\;\gamma\,g^{00}
\;+\;\partial_k(e_{a0}\Pi^{ak}) \;.\eqno(27)$$

\noindent The quantities $\alpha_{ik}$, $\beta_k$ and $\gamma$
are Lagrange multipliers.

Next we note that since the momenta $\{\Pi^{a0}\}$ are identically
vanishing, they also constitute primary constraints, which induce
the secondary constraints

$$C^a\; \equiv \;  {{\delta H}\over {\delta e_{a0}}}\;
=\;0\;.\eqno(28)$$

\noindent In the proccess of varying $H$ with respect to $e_{a0}$ 
we only have to consider $H_0$, because variation of the constraints
lead to the contraints themselves:

$${\delta \over {\delta e_{a0}}}\,(\Pi^{[ik]}-k\,e\,p^{ik})\;=\;
-{1\over 2}\biggl( e^{ai}(\Pi^{0k}-k\,e\,p^k)-
e^{ak}(\Pi^{0i}-k\,e\,p^i) \biggr)\;,\eqno(29)$$

$${\delta \over {\delta e_{a0}}}\,(\Pi^{0k}-k\,e\,p^k)\;=\;
-e^{a0}\,(\,\Pi^{0k}-k\,e\,p^k\,)\;.\eqno(30)$$

As we will see, the evaluation of the constraints $C^a$ 
according to (28) reveals the constraint structure of the 
Hamiltonian $H_0$. After a long calculation we arrive at

$$C^a\;=\;-\partial_k \Pi^{ak} + ke\,e^{a0}\biggl\{
{1\over 4} g^{im}g^{nj}T^b\,_{mn}T_{bij}+
{1\over 2}g^{jn}T^i\,_{mn}T^m\,_{ij}-
g^{ik}T^j\,_{ji}T^n\,_{nk}$$

$$+{1\over 2}g_{0i}(g_{0m}g_{nk}-{1\over 2}g_{0k}g_{mn})P^{mn}P^{ik}\biggr\}
-ke\,e^{ai}\biggl\{g^{0m}g^{nj}T^b\,_{ij}T_{bmn}$$

$$+g^{0j}T^m\,_{ni}T^n\,_{mj}+g^{nj}T^0\,_{mn}T^m\,_{ij}
-2g^{0k}T^j\,_{ji}T^n\,_{nk}-2g^{jk}T^0\,_{ij}T^n\,_{nk}\biggr\}$$

$$+ke\biggl\{
e^{aj}g_{ij}(g_{0m}g_{nk}-{1\over 2}g_{0k}g_{mn})P^{mn}P^{ik}$$

$$+{1\over 2}g_{0i}(g_{0m}g_{nk}-{1\over 2}g_{0k}g_{mn})
(P^{mn}\gamma^{aik}+P^{ik}\gamma^{amn})\biggr\}\;.\eqno(31)$$

\bigskip
\noindent The quantity $\gamma^{aik}$ appearing in (31) is defined
by

$$\gamma^{aik}\;=\;e^{al}\biggl\{g^{0j}(g^{0k}T^i\,_{lj}+
g^{0i}T^k\,_{lj})-2g^{0i}g^{0k}T^j\,_{lj}
+(g^{kj}g^{0i}+g^{ij}g^{0k}-2g^{ik}g^{0j})T^0\,_{lj}\biggr\}
\;.\eqno(32)$$

We immediately note that $C^a$ satisfies the relation

$$e_{a0}C^a\;=\;H_0\;.\eqno(33)$$

\noindent Therefore we can write the final form of the 
completely constrained Hamiltonian:

$$H\;=\;e_{a0}C^a\;+\;\alpha_{ik}\,(\Pi^{[ik]}-kep^{ik})\;+\;
\beta_k\,(\Pi^{0k}-kep^k)\;+\;\gamma\,g^{00}
\;+\;\partial_k(e_{a0}\Pi^{ak})\;.\eqno(34)$$

Although $e_{a0}$ appears as a Lagrange multiplier, it is also
contained  both in $H_0$ and in $C^a$. However, 
it is possible to check that $e_{a0}$ is a true Lagrange multiplier.
By just making use of the orthogonality relations of the tetrads,
it is possible to verify that the constraints $C^a$ satisfy the
relation

$$e_{a0}\,{{\delta C^a}\over {\delta e_{b0}}}\;=\;0\;,$$

\noindent from what we conclude that variation of $H$ given by (34)
with respect to $e_{a0}$ yields $C^a$ plus the constraints on the                     
right hand side of (29) and (30).     \\

\bigskip
\bigskip
\noindent {\bf V. Comments}\par
\bigskip
\noindent In the last section we have completed the 3+1 decomposition
of the Lagrangian density (4) on a null surface. In this
procedure all tetrad and metric components are four dimensional
quantities. We have not established any decomposition for these
fields, basically because it was not needed. Since the tetrads
do not obey any particular gauge condition,
the nondynamical component $e_{(0)0}$ cannot be identified with
the usual lapse function.

The final form of the Hamiltonian, eq.(34), is written as 
a sum of the constraints of the theory.
One major difference between this Hamiltonian formulation and the 
ADM-type  formulation is that in the latter the usual vector
constraint $H_i$ is linear in the momenta\cite{ADM}, whereas here both
$C^{(i)}$ and $C^{(0)}$ are linear and quadratic in $\Pi^{(ik)}$,
in general.

The next step is the determination of the constraint
algebra. The algebra of the ten constraints, equations (18), (19)
and (31), is expected to be quite intricate. The analysis of
\cite{Goldberg3,Goldberg4} showed the existence of second class
constraints. It is likely the same complication arises here. 
This issue will be investigated in the near future.

As we mentioned in the introduction, one major motivation for 
the present analysis is the establishment of an expression for the 
gravitational energy-momentum  vector density. In the present
case this expression is restricted to configurations of the
gravitational field that describe gravitational waves. Our 
previous experience on this subject lead us to conclude that
the covariant gravitational energy-momentum $P^a$ is given by

$$P^a\;=\;-\int_V d^3x \,\partial_k \Pi^{ak}\;.\eqno(35)$$

\noindent As before\cite{Maluf2}, the integral form of the
constraint $C^{(0)}=0$ can be interpreted as an energy
equation of the type $H - E=0$. Expression (35) allows us to
compute the energy-momentum of the gravitational radiation field
for an arbitrary volume of the three-dimensional space.
This analysis will be carried out in the context of the Bondi
and Sachs metrics and  presented in detail elsewhere.               \\

\bigskip
\bigskip
\noindent {\it Acknowledgements}\par
\noindent This work was supported in part by CNPQ. J. F. R. N.
is supported by CAPES, Brazil.\par

\end{document}